\documentclass[twocolumn,showpacs,preprintnumbers,amsmath,amssymb]{revtex4}

\usepackage{graphicx}
\usepackage{dcolumn}
\usepackage{bm}

\begin{document}

\title{Continuously decoupling single-qubit operations from a perturbing thermal bath of scalar bosons}

\author{F. F. Fanchini}
 \email{felipe@ifsc.usp.br}
 
\author{J. E. M. Hornos}

\author{R. d. J. Napolitano}
\affiliation{Instituto de F\'{\i}sica de S\~{a}o Carlos, Universidade de S\~{a}o Paulo, Caixa Postal 369, 13560-970, S\~{a}o Carlos, SP, Brazil}

\date{\today}

\begin{abstract}
We investigate the use of continuously-applied external fields to maximize the fidelity of quantum logic operations performed on a decohering qubit. Assuming a known error operator and an environment represented by a scalar boson field at a finite temperature, we show how decoherence during logical operations can be efficiently reduced by applying a superposition of two external vector fields: one rotating orthogonally to the direction of the other, which remains static. The required field directions, frequency of rotation and amplitudes to decouple noise dynamically are determined by the coupling constants and the desired logical operation. We illustrate these findings numerically for a Hadamard quantum gate and an environment with ohmic spectral density.
\end{abstract}

\pacs{03.67.Pp, 03.67.Lx, 03.67.-a, 03.65.Yz}

\maketitle

A quantum computer, when finally built, will be more efficient
than current classical computers to solve certain kinds of problems
\cite{deutsch92}. The theory of quantum information processing generally takes advantage of the inherent parallelism exhibited by unitary operations on quantum-state superpositions. The terms of these linear combinations are tensor products of quantum bits, or ``qubits'' \cite{schumacher95}, which, linearly superposed, result in states with the desired properties of entanglement and interference \cite{nielsen00}. In principle, the choice of an appropriate external field would guarantee a correct dynamics for the system, selected among those exhibiting unitary symmetry. However, during the actual quantum evolution of the system, since it cannot be completely separated from its environment, the unitary symmetry breaks down. The consequent decay of the quantum state purity is a manifestation of the ubiquitous phenomenon of decoherence \cite{zurek91}.

There are at least three major classes of strategic devices proposed to counteract the deleterious and unavoidable effects of decoherence: quantum error correcting codes \cite{shor95}, decoherence-free subspaces and subsystems \cite{zanardi97}, and dynamical decoupling \cite{viola98, viola99, viola99b,viola04}. Because the first two of these strategies require more than one physical qubit to protect each logical qubit, dynamical decoupling is the simplest of the three, since it requires, in principle, only controllable external fields to directly protect each physical qubit. Even without precise knowledge of the error structure and strengths, the pulsed dynamical-decoupling scheme is effective, but usually employ an articulate time sequence of external-field pulses which, for experimental implementations, requires sophisticated control procedures. Moreover, the pulses have to be so short as to start and finish well within the environmental correlation time interval, so the field intensities involved must be high. Initial attempts to use continuously-applied fields instead of pulses have appeared recently \cite{romero04} and these preliminary analyses show that, although pulses are not necessary for protecting against the effects of particular error structures assumed known, employing fast control cycles is inevitable. From the practical point of view of experimental realization, however, the successful introduction of continuous control fields into the general dynamical-decoupling framework will be relevant even in the high-frequency regime.

In this paper we show that, if the error structure and strengths are known, by continuously applying a suitable superposition of external fields it is possible to realize, with high fidelity, any logical operation on a qubit weakly perturbed by an environment represented by a scalar boson field at a finite temperature. The general formulation of the dynamical decoupling method \cite{viola99,facchi05} gives a clear geometrical interpretation of the error protection when explicitly applied to the simple case of a single qubit driven by continuous fields. Hence, under these circumstances, we have found that decoherence during logical operations can be efficiently reduced by applying a superposition of two external vector fields: one that rotates orthogonally to the direction of the other, which is a static vector field. The amplitudes, frequency, and directions of these fields are determined by the intended quantum logic operation, the error structure, and the characteristics of the environment.

For the purpose of starting an investigation on the use of continuous control fields to protect against errors during qubit operations, we make the simplifying assumption that the interaction between the qubit and its environment is sufficiently weak that perturbation theory is applicable. For arbitrary coupling strengths to undesirable but known terms in the qubit-system Hamiltonians, without taking into account environmental degrees of freedom that are to be traced out, there are the optimal-control approaches of Ref. \cite{zhang06}. In the present work we imagine a situation in which the qubit is far from other qubits, but is acted upon by controllable external fields and is subject to residual influences from the surround environment. For this particular setting, Ref. \cite{schulte06} presents an optimal control theory where the environmental perturbations are modeled by a Lindblad master equation, which is valid in the Markovian limit \cite{lindblad76}. Since here we treat the details of the controlling cycles occurring during the correlation time of the fields representing the environment, we must, although using perturbation theory, derive a non-Markovian master equation.

We start by noticing that the agents coupling the qubit to the environment can be represented by the Pauli matrices. For the sake of simplicity, we represent the action of the environment by a single scalar boson field and assume an interaction Hamiltonian given by
\begin{eqnarray}
H_{\rm int}= \left( {\bm \lambda} \cdot {\bm \sigma}\right)B+\left({\bm \lambda}^{\ast}\cdot {\bm \sigma}\right)B^{\dagger}, \label{Hint}
\end{eqnarray}
where ${\bm \lambda}$ is the error vector whose components are complex and we take $B=\sum_{k} g_{k}a_{k}$, where $g_{k}$ is a complex coupling constant for normal mode $k$ with dimension of frequency and $a_{k}$ is the operator that annihilates a bath quantum in mode $k$. Here and in the following we use units of $\hbar =1$.

Instead of starting from the Hamiltonian and deriving the propagator, which requires dealing with technical issues related to time ordering of interaction operators, we begin with the most general unitary evolution and obtain the Hamiltonian by differentiation. Any time-dependent unitary transformation of a qubit can always be expressed as
\begin{eqnarray}
U(t)=I\cos\left[\alpha(t)\right]-i {\bm \sigma}\cdot{\bf \hat{u}}(t)\sin\left[\alpha(t)\right],\label{U(t)}
\end{eqnarray}
where $I$ is the $2\times2$ identity matrix, $\alpha\left(t\right)$ is a function of time $t$, ${\bf \hat{u}}\left(t\right)$ is a time-dependent unit vector, ${\bm \sigma}={\bf \hat{x}}\sigma_{x}+{\bf \hat{y}}\sigma_{y}+{\bf \hat{z}}\sigma_{z}$, and $\sigma_{x}$, $\sigma_{y}$, and $\sigma_{z}$ are the Pauli matrices. The corresponding driving Hamiltonian $H_{U}(t)$ is obtained by differentiating Eq. (\ref{U(t)}):
\begin{eqnarray}
H_{U}(t)=i\frac{dU(t)}{dt}U^{\dagger}(t)={\bm \Omega}(t)\cdot {\bm \sigma},\label{HU}
\end{eqnarray}
with
\begin{eqnarray}
{\bm \Omega}(t)&=&\frac{d\alpha(t)}{dt}{\bf \hat{u}}(t)+\sin[\alpha (t)]\cos[\alpha (t)]\frac{d{\bf \hat{u}}(t)}{dt} \nonumber \\
&+&\sin ^{2}[\alpha (t)]{\bf \hat{u}}(t){\bf \times}\frac{d{\bf \hat{u}}(t)}{dt},\label{Omega}
\end{eqnarray}
where $U^{\dagger}(t)$ is the Hermitian conjugate of $U(t)$. It is important to emphasize that because of the results in Eqs. (\ref{U(t)})-(\ref{Omega}) we avoid any time-ordering manipulations and do not need the so-called polaron transformation \cite{jirari06}.

The time-local second-order master equation describing the evolution of the reduced density matrix of the qubit, in the interaction picture, is written as \cite{shibata77}
\begin{eqnarray}
\frac{d\rho _{I}(t)}{dt}=-\int^{t}_{0}dt^{\prime} {\rm Tr}_{B}\left\{{\left[H_{I}(t),\left[H_{I}(t^{\prime}),\rho _{B}\rho _{I}(t)\right]\right]}\right\},\label{master}
\end{eqnarray}
where $H_{I}(t)$ is the interaction Hamiltonian in the interaction picture, namely,
$H_{I}(t)=U^{\dagger}(t)U^{\dagger}_{B}(t)H_{\rm int}U_{B}(t)U(t)$, with
$U_{B}(t)=\exp\left(-iH_{B}t\right)$, $H_{B}=\sum _{k} \omega _{k} a^{\dagger}_{k}a_{k}$, $\omega _{k}$ is the frequency of normal mode $k$ of the thermal bath, and $U(t)$ is as in Eq. (\ref{U(t)}). Equation (\ref{master}) is valid in the regime in which the strength of the coupling, expressed in frequency units, multiplied by the correlation time of the bath operators is much lesser than unity.

Above, $\rho _{B}$ is the initial density matrix of the thermal bath:
\begin{eqnarray}
\rho _{B}=\frac{1}{Z}\exp(-\beta H_{B}),\label{rhoB}
\end{eqnarray}
where $Z$ is the partition function given by $Z={\rm Tr}_{B}\left[\exp(-\beta H_{B})\right]$. Here, $\beta =1/k_{B}T$, $k_{B}$ is Boltzmann constant, and $T$ is the absolute temperature of the environment.

From the form of the interaction between the qubit and its environment, Eq. (\ref{Hint}), we obtain
\begin{eqnarray}
H_{I}(t)= B_{I}(t){\bm \Lambda }(t)\cdot {\bm \sigma}+ B^{\dagger}_{I}(t){\bm \Lambda ^{\ast} }(t)\cdot {\bm \sigma},\label{HI3}
\end{eqnarray}
where $B_{I}(t)=U^{\dagger}_{B}(t)BU_{B}(t)=\sum_{k} g_{k}a_{k}\exp(-i\omega _{k}t)$ and we have defined the time-dependent vector ${\bm \Lambda }(t)$ as
\begin{eqnarray}
{\bm \Lambda }(t)&=&{\bm \lambda}\cos\left[2\alpha(t)\right]+\left[{\bm \lambda}{\bf \times}{\bf \hat{u}}(t)\right]\sin\left[2\alpha(t)\right]\nonumber\\
&+&{\bf \hat{u}}(t)\left[{\bf \hat{u}}(t)\cdot{\bm \lambda}\right]\left\{1-\cos\left[2\alpha(t)\right]\right\}.\label{Lambda}
\end{eqnarray}
Substituting Eq. (\ref{HI3}) into Eq. (\ref{master}), we obtain
\begin{eqnarray}
\frac{d\rho _{I}(t)}{dt}=\sum^{3}_{\alpha=1}\sum^{3}_{\beta=1} D_{\alpha \beta}(t)\left[\sigma _{\alpha},\rho _{I}(t)\sigma _{\beta}\right]\nonumber\\
+\sum^{3}_{\alpha=1}\sum^{3}_{\beta=1}D^{\ast}_{\alpha \beta}(t)\left[\sigma _{\beta}\rho _{I}(t),\sigma _{\alpha}\right],\label{master2}
\end{eqnarray}
where we have defined
\begin{eqnarray}
D_{\alpha \beta}(t)=\int^{t}_{0}dt^{\prime}{\rm Tr} _{B}\left[b_{\alpha }(t)\rho _{B}b_{\beta }(t^{\prime})\right],\label{D}
\end{eqnarray}
with the bath vector operator ${\bf b}(t)$ given by ${\bf b}(t)=B_{I}(t){\bm \Lambda }(t)+B^{\dagger}_{I}(t){\bm \Lambda ^{\ast} }(t)$.

Substituting (\ref{rhoB}) into (\ref{D}) gives
\begin{eqnarray}
D_{\alpha \beta}(t)=2{\rm Re}\left\{\Lambda ^{\ast}_{\alpha}(t)\int^{t}_{0}dt^{\prime}\Lambda _{\beta}(t^{\prime}){\cal I}_{1}(t-t^{\prime})\right\} \nonumber \\
+\Lambda ^{\ast}_{\alpha}(t)\int^{t}_{0}dt^{\prime}\Lambda _{\beta}(t^{\prime}){\cal I}_{2}(t-t^{\prime}),\label{D2}
\end{eqnarray}
where we have defined
\begin{eqnarray}
{\cal I}_{1}(t)&=&\sum _{k}\left|g_{k}\right|^{2}\frac{\exp\left(i\omega _{k}t\right)}{\exp(\beta  \omega _{k})-1}, \label{I1}\\
{\cal I}_{2}(t)&=&\sum _{k}\left|g_{k}\right|^{2}\exp\left(i\omega _{k}t\right). \label{I2}
\end{eqnarray}

In the limit in which the number of bath normal modes per unit frequency becomes infinite,
we define its spectral density as
\begin{eqnarray}
J(\omega)=\sum _{k}\left|g_{k}\right|^{2}\delta (\omega -\omega _{k}),\label{spectral}
\end{eqnarray}
with $\omega \in [0,+\infty)$ and interpret the summations in Eqs. (\ref{I1}) and (\ref{I2}) as integrals over $\omega$: ${\cal I}_{1}(t)=\int ^{\infty}_{0}d\omega J(\omega)\exp(i\omega t)/[\exp(\beta  \omega )-1]$ and ${\cal I}_{2}(t)=\int ^{\infty}_{0}d\omega J(\omega )\exp(i\omega t)$.

If we now write $\rho_{I}(t)=I/2+{\bf r}(t)\cdot{\bm \sigma}$, the Bloch vector ${\bf r}(t)$ is real and, from Eqs. (\ref{master2}) and (\ref{D2}), satisfy the differential equation
\begin{eqnarray}
\frac{d{\bf r}(t)}{dt}&=&4{\rm Re}\left\{{\bm\Lambda}^{\ast}(t)\times\left[\left(2{\bf F}(t)+{\bf G}(t)\right)\times{\bf r}(t)\right]\right\}\nonumber\\
&&-2{\rm Im}\left[{\bm \Lambda}^{\ast}(t)\times{\bf G}(t)\right],\label{Bloch}
\end{eqnarray}
where we have defined ${\bf F}(t)=\int_{0}^{t}dt^{\prime}{\bm \Lambda}(t^{\prime}) {\cal I}_{1}(t-t^{\prime})$ and ${\bf G}(t)=\int_{0}^{t}dt^{\prime}{\bm \Lambda}(t^{\prime}){\cal I}_{2}(t-t^{\prime})$.

The vectors ${\bf F}(t)$ and ${\bf G}(t)$ can be interpreted as time averages of the vector ${\bm \Lambda}(t^{\prime})$ weighted by the functions ${\cal I}_{1}(t-t^{\prime})$ and ${\cal I}_{2}(t-t^{\prime})$, respectively. Depending on the width of the reservoir spectral density, $J(\omega)$ in Eq. (\ref{spectral}), the weight functions, ${\cal I}_{1}(t-t^{\prime})$ and ${\cal I}_{2}(t-t^{\prime})$, determine how much of the recent past history of ${\bm \Lambda}(t^{\prime})$ effectively contributes to its time averages, ${\bf F}(t)$ and ${\bf G}(t)$. Hence, Eq. (\ref{Bloch}) is not restricted by a Markovian approximation.

We aim at finding $U(t)$ such that Eq. (\ref{Bloch}), which refers to the interaction picture, gives, after a certain gate-operation time $\tau$, ${\bf r}(\tau)\approx{\bf r}(0)$. In the Schr\"{o}dinger picture, this result means that, effectively, the dynamics described by Eq. (\ref{Bloch}) is equivalent to the action of $U(\tau)$, as if the qubit were not perturbed by its environment. The total Hamiltonian, in the Shr\"{o}dinger picture, is given by $H(t)=H_{U}(t)+H_{B}+H_{\rm int}$, where $H_{U}(t)$ is to be determined as prescribed by Eq. (\ref{HU}) after we find $U(t)$, $H_{B}$ is the bath Hamiltonian as described just below Eq. (\ref{master}), and $H_{\rm int}$ is the interaction Hamiltonian as given by Eq. (\ref{Hint}). 

Because we intend to realize a quantum logic operation simultaneously to the protection from errors, we split $H_{U}$ into two terms, $H_{U}(t)=H_{0}(t)+H_{c}(t)$, where $H_{0}(t)$ is to produce the quantum-gate action, while $H_{c}(t)$ counteracts the perturbing action of the environment. According to the prescription well exposed in Ref. \cite{facchi05}, the unitary operator $U_{c}(t)$ corresponding to the Hamiltonian $H_{c}(t)$ is to be periodic and satisfy $\int_{0}^{t_{c}}dtU_{c}^{\dagger}(t)H_{\rm int}U_{c}(t)=0$, where $t_{c}<\tau$ is the period of $U_{c}(t)$, that is, $U_{c}(t+t_{c})=U_{c}(t)$. Here, for convenience, we choose $\tau$ as an integer multiple $n$ of $t_{c}$, that is, $\tau=nt_{c}$, such that $U_{c}(\tau)=I$. If we choose a constant $H_{c}=(2\pi/t_{c}){\bf \hat{u}}_{c}\cdot{\bm \sigma}$, then $U_{c}(t)=I\cos(2n\pi t/\tau)-i{\bf \hat{u}}_{c}\cdot{\bm \sigma}\sin(2n\pi t/\tau)$. By the same procedure to obtain Eqs. (\ref{HI3}) and (\ref{Lambda}), we find that the integral of $U_{c}^{\dagger}(t)H_{\rm int}U_{c}(t)$ over one period $t_{c}$ gives zero only if we choose ${\bf \hat{u}}_{c}$ orthogonal to the error vector ${\bm \lambda}$, which is always possible, even if ${\bm \lambda}$ has complex components.

In the picture obtained using the unitary transformation $U_{c}(t)$ as given above, we choose $H_{0}(t)$ such that $U_{c}^{\dagger}(t)H_{0}(t)U_{c}(t)=I\cos\theta _{0}-i {\bm \sigma}\cdot{\bf \hat{u}_{0}}\sin\theta _{0}$, where ${\bf \hat{u}}_{0}$ and $\theta _{0}$ are constants to be determined according to the intended quantum logic operation as explained in the following. Since $U_{c}^{\dagger}(t)H_{0}(t)U_{c}(t)$ is time independent, the associated unitary evolution operator is given by $U_{0}(t)=I\cos(\theta _{0} t/\tau)-i{\bf \hat{u}}_{0}\cdot{\bm \sigma}\sin(\theta _{0} t/\tau)$, giving $U_{0}(\tau)=I\cos\theta _{0}-i {\bm \sigma}\cdot{\bf \hat{u}_{0}}\sin\theta _{0}$ at the end of the gate operation, $t=\tau$. Therefore, if we take $U(t)$ as the composed unitary operator $U(t)=U_{c}(t)U_{0}(t)$, then, at $t=\tau$, $U(\tau)=I\cos\theta _{0}-i {\bm \sigma}\cdot{\bf \hat{u}_{0}}\sin\theta _{0}$, so that the intended operation determines our choice of ${\bf \hat{u}}_{0}$ and $\theta _{0}$. Comparing these conclusions with Eq. (\ref{U(t)}), we obtain:
\begin{eqnarray}
\cos[\alpha(t)]&=&-{\bf \hat{u}}_{c}\cdot{\bf \hat{u}}_{0}\sin(2n\pi t/\tau)\sin(\theta _{0}/\tau)\nonumber\\
& &+\cos(2n\pi t/\tau)\cos(\theta _{0}/\tau),\label{cos}\\
{\bf \hat{u}}(t)\sin[\alpha(t)]&=&({\bf \hat{u}}_{c}\times{\bf \hat{u}}_{0})\sin(2n\pi t/\tau)\sin(\theta _{0}/\tau)\nonumber\\
& &+{\bf \hat{u}}_{c}\sin(2n\pi t/\tau)\cos(\theta _{0}/\tau)\nonumber\\
& &+{\bf \hat{u}}_{0}\cos(2n\pi t/\tau)\sin(\theta _{0}/\tau)\label{usin}.
\end{eqnarray}

The explicit form of the Hamiltonian $H_{U}$ is the one already given by Eq. (\ref{HU}), $H_{U}={\bm \Omega}(t)\cdot {\bm \sigma}$, where the applied external field is calculated from Eqs. (\ref{cos}) and (\ref{usin}) according to the prescription of Eq. (\ref{Omega}):
\begin{eqnarray}
{\bm \Omega}(t)&=&[(2n\pi/\tau)+(\theta _{0}/\tau){\bf \hat{u}}_{c}\cdot{\bf \hat{u}}_{0}]{\bf \hat{u}}_{c}\nonumber\\
& &+(\theta _{0}/\tau)[{\bf \hat{u}}_{c}\times({\bf \hat{u}}_{0}\times{\bf \hat{u}}_{c})]\cos(2n\pi t/\tau)\nonumber\\
& &+(\theta _{0}/\tau)({\bf \hat{u}}_{c}\times{\bf \hat{u}}_{0})\sin(2n\pi t/\tau).\label{field}
\end{eqnarray}
The first term of Eq. (\ref{field}) is a static field along the direction that is perpendicular to the error vector, as discussed above, and the other two terms give a rotating field perpendicular to the direction of the static field.

\begin{figure}
\includegraphics[width=9cm]{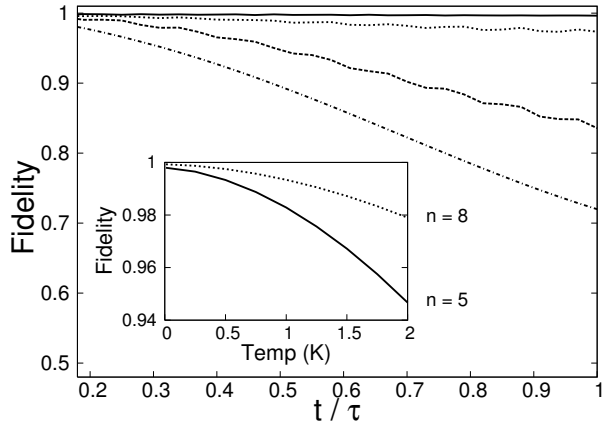}
\caption{\label{figure1} Numerical solutions for the strategy example described in the text to overcome decoherence and dissipation during a Hadamard operation. The dot-dashed line represents the fidelity when $H_{c}=0$, while the solid line is the result for $n=5$ in Eq. (\ref{field}) and ${\bf \hat{u}}_{c}$, chosen orthogonally to the error vector ${\bm \lambda}$, as in the text. The dotted and dashed lines represent the $n=5$ results when ${\bf \hat{u}}_{c}$ is tilted $10^{\circ}$ and $30^{\circ}$ toward the $x$ axis, respectively. In the inset we show the fidelities at $t=\tau$ as functions of temperature and $n$.}
\end{figure}

We have performed numerical calculations for the strategy described above by solving Eq. (\ref{Bloch}). To illustrate our typical results for a concrete example, in the present article we assume an ohmic spectral density with a cutoff frequency $\omega _{c}$, namely, $J(\omega )=\eta \omega \exp(-\omega / \omega _{c})$, where $\eta $ is a dimensionless constant. Hence, the integral versions of Eqs. (\ref{I1}) and (\ref{I2}) can be explicitly calculated to give
${\cal I}_{1}(t)=(\eta /\beta ^{2}) \Psi ^{(1)}\left(1+1/(\beta  \omega _{c})-it/\beta \right)$, and ${\cal I}_{2}(t)=\eta \omega _{c}^{2}/\left(1-i\omega _{c}t\right)^{2}$, where $\Psi ^{(1)}$ is the first polygamma function. Since, as in the case of pulsed dynamical decoupling \cite{viola98,viola99,viola99b}, we have found that the rotating field, to be effective, must rotate at a frequency sufficiently higher than $\omega_{c}$, we take $\omega_{c}\tau=2\pi$ in the numerical calculations. One of the worst situations occurs when the error vector has complex components, describing, besides pure decoherence, also dissipation. Thus, we take ${\bm \lambda}=(4{\bf \hat{x}}+i{\bf \hat{y}}+2\sqrt{2}{\bf \hat{z}})/5$, since it is dimensionless according to Eq. (\ref{Hint}). Thus, ${\bf \hat{u}}_{c}=({\bf \hat{x}}-\sqrt{2}{\bf \hat{z}})/\sqrt{3}$ is a suitable choice. For a Hadamard operation, we can choose ${\bf \hat{u}}_{0}=({\bf \hat{x}}+{\bf \hat{z}})/\sqrt{2}$ \cite{nielsen00}. We take $\eta=1/16$, $T=0.25$K, and $\tau=10^{-10}$s. With these numbers and the initial condition $\rho_{I}(0)=I/2+\sigma_{x}/2$, when $H_{c}=0$ the fidelity gives ${\cal F}(\tau)={\rm Tr}[\rho_{I}(\tau)\rho_{I}(0)]\approx0.7199$. Now, if we take $n=5$ in Eq. (\ref{field}), the fidelity becomes ${\cal F}(\tau)\approx0.9965$. Figure $1$ shows these fidelities as functions of time, together with two cases in which ${\bf \hat{u}}_{c}$ is not chosen as orthogonal to the error vector, but tilted $10^{\circ}$ and $30^{\circ}$ toward the $x$ axis. In these examples we see that the resulting fidelity function is only smoothly sensitive to small-angle variations. Incidentally, since the condition for the present scheme to work depends only on the angle between ${\bf \hat{u}}_{c}$ and ${\bm \lambda}$, the error protection is also smoothly dependent on the error-vector direction. In the inset, we show the final fidelities, calculated at $t=\tau$, as functions of temperature and $n$ with ${\bf \hat{u}}_{c}$ as given above, orthogonal to ${\bm \lambda}$; we see that lower temperatures are better for increasing fidelity, but higher-temperature effects can be compensated by higher control-field frequencies.

The initial conditions are specified by ${\bf r}(0)$, whose modulus, for pure initial states, is $1/2$. Thus, all the initial conditions for pure initial states can be parametrized by two angles: $\varphi \in [0,2\pi)$ and $\theta \in [0,\pi]$. We have partitioned these angle ranges into $200$ and $100$ regularly-spaced intervals, respectively, and have solved Eq. (\ref{Bloch}) for each of these $20,000$ initial conditions. The worst and best cases resulted in fidelities of $0.99622$ and $0.99987$, respectively, with all the other variables set as for the solid line in Fig. $1$.

In summary, we have shown that it is possible, using continuously-applied external fields, to realize one-qubit high-fidelity quantum logic operations and, simultaneously, protect these operations against environmental errors. Our proposed continuous dynamical decoupling is less vulnerable to errors produced by non-ideal driving than the pulsed version: our method requires intervention only at the initial and final steps of the quantum logic operation, while the pulsed method demands sophisticated timing control for each of the various necessary pulses. 

\begin{acknowledgments}
This work has been supported by Funda\c{c}\~{a}o de Amparo \`{a} Pesquisa do Estado de S\~{a}o
Paulo, Brazil, project number 05/04105-5 and the Millennium Institute for Quantum Information -- Conselho Nacional de Desenvolvimento Cient\'{\i}fico e Tecnol\'{o}gico, Brazil. We also acknowledge careful reading and comments by Drs. M. A. Marchiolli and P. E. M. F. de Mendon\c{c}a.
\end{acknowledgments}

\end{document}